# On the Complexity of Local Search for Weighted Standard Set Problems[*]

Dominic Dumrauf[1,2] and Tim Süß[2]

[1] Paderborn Institute for Scientific Computation, 33102 Paderborn, Germany.
`dumrauf@uni-paderborn.de`

[2] Institute of Computer Science, University of Paderborn, 33102 Paderborn, Germany.
`tsuess@uni-paderborn.de`

**Abstract.** In this paper, we study the complexity of computing *locally optimal* solutions for *weighted* versions of *standard set problems* such as SETCOVER, SETPACKING, and many more. For our investigation, we use the framework of $\mathcal{PLS}$, as defined in Johnson et al., [13]. We show that for most of these problems, computing a locally optimal solution is already $\mathcal{PLS}$-complete for a simple natural neighborhood of size one. For the local search versions of weighted SETPACKING and SETCOVER, we derive *tight bounds* for a simple neighborhood of size two. To the best of our knowledge, these are one of the very few $\mathcal{PLS}$ results about local search for weighted standard set problems.

## 1 Introduction

*Set Problems and Their Approximation* In this paper, we study the complexity of computing *locally optimal* solutions for *weighted standard set problems* in the framework of $\mathcal{PLS}$, as defined in Johnson et al., [13]. In weighted set problems such as SETPACKING or SETCOVER, the input consists of a *set system* along with a *weight function* on the set system. The task is to compute a solution maximizing or minimizing some objective function on the set system while obeying certain constraints. Weighted set problems are *fundamental combinatorial optimization* problems with a wide range of applications spanning from *crew scheduling* in transportation networks and *machine scheduling* to *facility location* problems. Since they are of such fundamental importance on the one hand but of *computational intractability* on the other hand, [8], the *approximation* of weighted standard set problems has been extensively studied in the literature. Numerous heuristics have been applied to or developed for these problems, spanning from greedy algorithms and linear programming to local search.

*Local Search and Set Problems* *Local search* is a standard approach to approximate solutions of hard combinatorial optimization problems. Starting from an arbitrary feasible solution, a sequence of feasible solutions is iteratively generated, such that each solution is contained in the predefined *neighborhood* of its predecessor solution and strictly improves a given *cost function*. If no improvement within the neighborhood of a solution is possible, a *local optimum* (or *locally optimal solution*) is found. In practice, local search algorithms often require only a few steps to compute a solution. However, the running time is often pseudo-polynomial and even exponential in the worst case.

[*] This work has been partially supported by German Research Foundation (DFG) Priority Programm 1307 Algorithm Engineering, DFG-Project DA155/31-1, ME872/11-1, FI14911-1 AVIPASIA and the DFG Research Training Group GK-693 of the Paderborn Institute for Scientific Computation. Some of the work was done while the first author visited Simon Fraser University.

***Polynomial Time Local Search*** Johnson, Papadimitriou, and Yannakakis, [13], introduced the class $\mathcal{PLS}$ (polynomial-time local search) in 1988 to investigate the complexity of local search algorithms. Essentially, a problem in $\mathcal{PLS}$ is given by some *minimization* or *maximization problem* over instances with finite sets of feasible solutions together with a non-negative *cost function*. A *neighborhood structure* is superimposed over the set of feasible solutions, with the property that a local improvement in the neighborhood can be found in polynomial time. The objective is to find a *locally optimal* solution. The notion of a $\mathcal{PLS}$-reduction was defined in Johnson et al., [13], to establish relationships between $\mathcal{PLS}$-problems and to further classify them. Not many problems are known to be $\mathcal{PLS}$-complete, since reductions are mostly technically involved, which seems to be in large parts due to the transformation of the neighborhood under the reduction. In the recent past, *game theoretic approaches* re-raised the focus on the class $\mathcal{PLS}$ since in many games the computation of a Nash Equilibrium can be modeled as a local search problem, [7]. The knowledge about $\mathcal{PLS}$ is still very limited and not at all comparable with our rich knowledge about $\mathcal{NP}$.

In this paper, we show that for most weighted standard set problems, computing locally optimal solutions is $\mathcal{PLS}$-complete, even for very *small natural neighborhoods*. This implies that computing local optima for these problems via successive improvements may not yield a sufficient performance improvement over computing globally optimal solutions. Furthermore, we believe that most problems investigated in this paper have the potential to serve as candidates for the base of future reductions.

## 2  Notation and Contribution

In this section, we describe the notation, complexity classes, and problems considered throughout this paper. The fundamental definitions of a $\mathcal{PLS}$-problem and the class $\mathcal{PLS}$ were introduced by Johnson, Papadimitriou, and Yannakakis, [13]. For all $k \in \mathbb{N}$, denote $[k] := \{1, \ldots, k\}$, and $[k]_0 := [k] \cup \{0\}$. Given a $k$-tuple $T$, let $P_i(T)$ denote the projection to the $i$-th coordinate for some $i \in [k]$. For some set $S$, denote by $2^{|S|}$ the power set of $S$.

***$\mathcal{PLS}$, Reductions, and Completeness, [13]*** A $\mathcal{PLS}$-problem $L = (D_L, F_L, c_L, N_L, \text{INIT}_L, \text{COST}_L, \text{IMPROVE}_L)$ is characterized by seven parameters. The set of instances is given by $D_L \subseteq \{0,1\}^*$. Every instance $I \in D_L$ has a set of feasible solutions $F_L(I)$, where feasible solutions $\mathsf{s} \in F_L(I)$ have length bounded by a polynomial in the length of $I$. Every feasible solution $\mathsf{s} \in F_L(I)$ has a non-negative integer cost $c_L(\mathsf{s}, I)$ and a neighborhood $N_L(\mathsf{s}, I) \subseteq F_L(I)$. $\text{INIT}_L(I)$, $\text{COST}_L(\mathsf{s}, I)$, and $\text{IMPROVE}_L(\mathsf{s}, I)$ are polynomial time algorithms. Algorithm $\text{INIT}_L(I)$, given an instance $I \in D_L$, computes an initial feasible solution $\mathsf{s} \in F_L(I)$. Algorithm $\text{COST}_L(\mathsf{s}, I)$, given a solution $\mathsf{s} \in F_L(I)$ and an instance $I \in D_L$, computes the cost of the solution. Algorithm $\text{IMPROVE}_L(\mathsf{s}, I)$, given a solution $\mathsf{s} \in F_L(I)$ and an instance $I \in D_L$, finds a better solution in $N_L(\mathsf{s}, I)$ or returns that there is no better one.

A solution $\mathsf{s} \in F_L(I)$ is *locally optimal*, if for every neighboring solution $\mathsf{s}' \in N_L(\mathsf{s}, I)$ it holds that $c_L(\mathsf{s}', I) \leq c_L(\mathsf{s}, I)$ in case $L$ is a maximization $\mathcal{PLS}$-problem and $c_L(\mathsf{s}', I) \geq c_L(\mathsf{s}, I)$ in case $L$ is a minimization $\mathcal{PLS}$-problem. A *search problem* $R$ is given by a relation over $\{0,1\}^* \times \{0,1\}^*$. An algorithm "solves" $R$, when given $I \in \{0,1\}^*$ it computes an $\mathsf{s} \in \{0,1\}^*$, such that $(I, \mathsf{s}) \in R$ or it correctly outputs that such an $\mathsf{s}$ does not exist. Given a $\mathcal{PLS}$-problem $L$, let the according search problem be $R_L := \{(I, \mathsf{s}) \mid I \in D_L, \mathsf{s} \in F_L(I) \text{ is a local optimum}\}$. The *class* $\mathcal{PLS}$ is defined as $\mathcal{PLS} := \{R_L \mid L \text{ is a } \mathcal{PLS}\text{-problem}\}$. A $\mathcal{PLS}$-problem $L_1$ is $\mathcal{PLS}$-reducible to a $\mathcal{PLS}$-problem $L_2$ (written $L_1 \leq_{\text{pls}} L_2$), if there exist two polynomial-time computable functions $\Phi : D_{L_1} \mapsto D_{L_2}$ and $\Psi$ defined for $\{(I, \mathsf{s}) \mid I \in D_{L_1}, \mathsf{s} \in F_{L_2}(\Phi(I))\}$ with $\Psi(I, \mathsf{s}) \in F_{L_1}(I)$, such that for all $I \in D_{L_1}$ and for all $\mathsf{s} \in F_{L_2}(\Phi(I))$ it holds that, if $(\Phi(I), \mathsf{s}) \in R_{L_2}$, then $(I, \Psi(I, \mathsf{s})) \in R_{L_1}$. A $\mathcal{PLS}$-problem $L$ is $\mathcal{PLS}$-complete if every $\mathcal{PLS}$-problem is $\mathcal{PLS}$-reducible to $L$.

We write limitations to a problem as a prefix and the size of the neighborhood as a suffix. For all $\mathcal{PLS}$-problems $L$ studied in this paper, the algorithms $\text{INIT}_L$, $\text{COST}_L$, and $\text{IMPROVE}_L$ are straightforward and polynomial-time computable.

## 2.1 Weighted Standard Set Problems

We next describe the $\mathcal{PLS}$-problems we study in this paper. All problems we present are local search versions of their respective decision problems. In the following, let $\mathcal{B}$ denote some finite set and let $\mathcal{C} := \{C_1, \ldots, C_n\}$ denote a collection of subsets over $\mathcal{B}$. Let $w_\mathcal{C} : \mathcal{C} \mapsto \mathbb{N}$ and $w_\mathcal{B} : \mathcal{B} \times \mathcal{B} \mapsto \mathbb{N}$. Denote by $m_\mathcal{B}$ and $m_\mathcal{C}$ positive integers with $m_\mathcal{B} \leq |\mathcal{B}|$ and $m_\mathcal{C} \leq |\mathcal{C}|$. Unless otherwise mentioned, we use the *k-differ-neighborhood* where two solutions are mutual neighbors if they differ in at most $k$-elements which describe a solution. Except for SETCOVER, all problems are maximization problems.

**Definition 1 (W3DM-(p,q), [6]).** *$I \in D_{\text{W3DM}}$ of WEIGHTED-3-DIMENSIONALMATCHING (in short W3DM) is a pair $(n, w)$ with $n \in \mathbb{N}$ and $w$ is a function $w : [n]^3 \to \mathbb{R}_{\geq 0}$. The components of triples are identified with boys, girls, and homes. $F_{\text{W3DM}}(I)$ are all matchings of boys, girls, and homes, i.e. all $\mathsf{S} \subseteq [n]^3$, with $|\mathsf{S}| = n$, $P_k(T_i) \neq P_k(T_j)$, for all $T_i, T_j \in \mathsf{S}$, $i \neq j$, $k \in [3]$. For $\mathsf{S} \in F_{\text{W3DM}}(I)$ the cost is $c_{\text{W3DM}}(\mathsf{S}, I) := \sum_{T_i \in \mathsf{S}} w(T_i)$. $N_{\text{W3DM-(p,q)}}(\mathsf{S}, I)$ contains all feasible solutions where at most $p$ triples are replaced and up to $q$ boys or girls move to new homes.*

**Definition 2 (X3C-(k)).** *$I \in D_{\text{X3C}}$ of EXACT-COVER-BY-3-SETS (in short X3C) is a collection $\mathcal{C} := \{C_1, \ldots, C_n\}$ of all 3-element sets over a finite set $\mathcal{B}$, with $|\mathcal{B}| = 3q$ for some $q \in \mathbb{N}$, and $w : \mathcal{C} \mapsto \mathbb{N}$. $F_{\text{X3C}}(I)$ are all $\mathsf{S} \subseteq \mathcal{C}$ such that every $b \in \mathcal{B}$ is in exactly one $C_i \in \mathsf{S}$. For $\mathsf{S} \in F_{\text{X3C}}(I)$ the cost is $c_{\text{X3C}}(\mathsf{S}, I) := \sum_{C_i \in \mathsf{S}} w(C_i)$.*

**Definition 3 (SP-(k)).** *$I \in D_{\text{SP}}$ of SETPACKING (in short SP) is a triple $(\mathcal{C}, w_\mathcal{C}, m_\mathcal{C})$. $F_{\text{SP}}(I)$ are all sets $\mathsf{S} \subseteq \mathcal{C}$ with $|\mathsf{S}| \leq m_\mathcal{C}$. For $\mathsf{S} \in F_{\text{SP}}(I)$ the cost is $c_{\text{SP}}(\mathsf{S}, I) := \sum_{C_i \in \mathsf{S} \wedge \forall j \in [m], j \neq i : C_i \cap C_j = \emptyset} w(C_i)$.*

**Definition 4 (SSP-(k)).** *$I \in D_{\text{SSP}}$ of SETSPLITTING (in short SSP) is a tuple $(\mathcal{C}, w_\mathcal{C})$. Feasible solutions $F_{\text{SSP}}(I)$ are all partitionings $S_1, S_2 \subseteq \mathcal{B}$ of $\mathcal{B}$. For $\mathsf{S} \in F_{\text{SSP}}(I)$ the cost is $c_{\text{SSP}}(\mathsf{S}, I) := \sum_{C_i \in \mathcal{C} \wedge \exists s_1 \in C_i : s_1 \in S_1 \wedge \exists s_2 \in C_i : s_2 \in S_2} w(C_i)$.*

**Definition 5 (SC-(k)).** *$I \in D_{\text{SC}}$ of SETCOVER (in short SC) is a tuple $(\mathcal{C}, w_\mathcal{C})$. $F_{\text{SC}}(I)$ are all subsets $\mathsf{S} \subseteq \mathcal{C}$ with $\bigcup_{C_i \in \mathsf{S}} C_i = \mathcal{B}$. For $\mathsf{S} \in F_{\text{SC}}(I)$ the cost is $c_{\text{SC}}(\mathsf{S}, I) := \sum_{C_i \in \mathsf{S}} w(C_i)$.*

**Definition 6 (TS-(k)).** *$I \in D_{\text{TS}}$ of TESTSET (in short TS) is a triple $(\mathcal{C}, w_\mathcal{B}, m_\mathcal{B})$. Feasible solutions $F_{\text{TS}}(I)$ are all sets $\mathsf{S} \subseteq \mathcal{C}$ with $|\mathsf{S}| \in [m_\mathcal{B}]$. For $\mathsf{S} \in F_{\text{TS}}(I)$ the cost is $c_{\text{TS}}(\mathsf{S}, I) := \sum_{b_i, b_j \in \mathcal{B}; i < j \wedge \exists C_p, C_q \in \mathsf{S} \text{ containing exactly one of } b_i \text{ and } b_j} w(b_i, b_j)$.*

**Definition 7 (SB-(k)).** *$I \in D_{\text{SB}}$ of SETBASIS (in short SB) is a triple $(\mathcal{C}, w_\mathcal{C}, m_\mathcal{C})$. $F_{\text{SB}}(I)$ are all sets $\mathsf{S} = \{S_1, \ldots, S_{m_\mathcal{C}}\}$, where $S_i \in 2^{|\mathcal{B}|}$ for all $i \in [m_\mathcal{C}]$. For $\mathsf{S} \in F_{\text{SB}}(I)$ the cost is $c_{\text{SB}}(\mathsf{S}, I) := \sum_{C_i \in \mathcal{C} \wedge \exists S' \subseteq \mathsf{S} : C_i = \bigcup_{S'_i \in S'} S'_i} w(C_i)$.*

**Definition 8 (HS-(k)).** *$I \in D_{\text{HS}}$ of HITTINGSET (in short HS) is a triple $(\mathcal{C}, w_\mathcal{C}, m_\mathcal{B})$. $F_{\text{HS}}(I)$ are all sets $\mathsf{S} \subseteq \mathcal{B}$ with $|\mathsf{S}| \leq m_\mathcal{B}$. For $\mathsf{S} \in F_{\text{HS}}(I)$ the cost is $c_{\text{HS}}(\mathsf{S}, I) := \sum_{C_i \in \mathcal{C} \wedge \exists s \in \mathsf{S} : s \in C_i} w(C_i)$.*

**Definition 9 (IP-(k)).** *$I \in D_{\text{IP}}$ of INTERSECTIONPATTERN (in short IP) are two symmetric $n \times n$ matrices $A = (a_{ij})_{i,j \in [n]}$ and $B = (b_{ij})_{i,j \in [n]}$ with positive integer entries and a collection $\mathcal{D} := \{D_1, \ldots, D_l\}$ with $l \geq n$ over a set $\mathcal{B}$. $F_{\text{IP}}(I)$ are all vectors $\mathcal{C} := (C_1, \ldots, C_n)$ with $C_i \in \mathcal{D}$ for all $i \in [n]$. For $\mathsf{S} \in F_{\text{IP}}(I)$ the cost is $c_{\text{IP}}(\mathsf{S}, I) := \sum_{i \leq j \in [n], |C_i \cap C_j| = a_{ij}} b_{ij}$.*

**Definition 10 (CC-(k)).** $I \in D_{\text{CC}}$ of COMPARATIVECONTAINMENT *(in short* CC*) are two collections* $\mathcal{C} := \{C_1, \ldots, C_n\}$, *and* $\mathcal{D} := \{D_1, \ldots, D_l\}$ *of sets over a set* $\mathcal{B}$, *and a function* $w : \mathcal{C} \cup \mathcal{D} \mapsto \mathbb{N}$. $F_{\text{CC}}(I)$ *are all sets* $\mathsf{S} \subseteq \mathcal{B}$. *For* $\mathsf{S} \in F_{\text{CC}}(I)$ *the cost is* $c_{\text{CC}}(\mathsf{S}, I) := \sum_{C_i \in \mathcal{C}; \mathsf{S} \subseteq C_i} w(C_i) - \sum_{D_i \in \mathcal{D}; \mathsf{S} \subseteq D_i} w(D_i) + W$, *where* $W \geq \sum_{D_i \in \mathcal{D}} w(D_i)$.

## 2.2 Generalized Satisfiability Problems

The hardness results we present in this paper, rely on known hardness results for the problems given below. For all these problems, we use the neighborhood where the value of one variable is changed. The task is to compute an assignment maximizing the sum of the weights.

**Definition 11 ((p,q,r)-MCA, [5]).** *An instance* $I \in D_{(p,q,r)\text{-MCA}}$ *of* $(p, q, r)$-MAXCONSTRAINT-ASSIGNMENT *is a set of constraints* $\mathcal{C} := \{C_1, \ldots, C_m\}$ *over a set of variables* $\mathcal{X} := \{x_1, \ldots, x_n\}$. *Every constraint* $C_i(x_{i_1}, \ldots, x_{i_{p_i}}) \in \mathcal{C}$ *has length at most* $p$ *and is a function* $w_{C_i} : [r]^{p_i} \mapsto \mathbb{R}_{\geq 0}$. *Every variable appears in at most* $q$ *constraints and takes values from* $[r]$ *with* $r \in \mathbb{N}$. $F_{\text{MCA}}(I)$ *are all assignments* $\mathsf{a} : \mathcal{X} \mapsto [r]$. *The cost of* $\mathsf{a} \in F_{\text{MCA}}(I)$ *is* $c_{\text{MCA}}(\mathsf{a}, I) := \sum_{C_i(x_{i_1}, \ldots, x_{i_{p_i}}) \in \mathcal{C}} w_{C_i}(\mathsf{a}(x_{i_1}), \ldots, \mathsf{a}(x_{i_{p_i}}))$.

**Definition 12 (POSNAE, [17]).** *An instance* $I \in D_{\text{POSNAE}}$ *of* POSITIVENOTALLEQUAL *(in short* POSNAE*) is an instance of* $(2, *, 2)$-MCA. *Constraints have length two and return the weight* $w_{C_i}$ *of constraint* $C_i \in \mathcal{C}$ *if the two literals in the clause do not have the identical assignment, otherwise they return 0.*

**Definition 13 ((h)-CNFSAT, [14]).** *An instance* $I \in D_{\text{CNFSAT}}$ *of* CNFSAT *is an instance of* $(h, *, 2)$-MCA. *Constraints are limited to disjunctions of literals over binary variables* $x \in \mathcal{X}$. *We drop the prefix if we refer to instances where clauses can have arbitrary length.*

## 2.3 Related Work

In this subsection, we mainly present related work about $\mathcal{PLS}$ and $\mathcal{PLS}$-completeness. The approximation of set problems has been intensively studied in the literature, [10–12, 16]. Survey articles about local search algorithms can be found in several books, [1, 2]. Local search for set problems has been applied in numerous papers, [4, 9]. For a survey on the quality of solutions obtained via local search not only for set problems, confer [3]. $\mathcal{PLS}$ was defined in Johnson et al., [13], and the fundamental definitions and results are presented in [13, 17]. Krentel, [14], shows that $(h)$-CNFSAT is $\mathcal{PLS}$-complete for some constant $h \in \mathbb{N}$. Schäffer and Yannakakis, [17], show that POSNAE, among numerous other local search problems, is $\mathcal{PLS}$-complete. The problem $(p, q, r)$-MCA is known to be $\mathcal{PLS}$-complete for triples (3,2,3), (2,3,8), and (6,2,2), [5, 14]. Orlin, Punnen, and Schulz present an FPTAS for computing approximate local optima for every linear combinatorial optimization problem in $\mathcal{PLS}$, [15]. The book of Aarts et al., [1], contains a list of $\mathcal{PLS}$-complete problems known so far.

## 2.4 Our Contribution

In this paper, we show that for most of the weighted standard set problems given in Subsection 2.1, computing a locally optimal solution is $\mathcal{PLS}$-*complete* for the *1-differ-neighborhood*. This means, that the problems are already hard, when one element describing the solution is allowed to be added, deleted, or exchanged for another element which is not part of the solution. As our main result, we prove the following two theorems:

**Theorem 1.** *The problems* SSP-*(k),* TS-*(k),* HS-*(k),* SB-*(k),* IP-*(k), and* CC-*(k) are $\mathcal{PLS}$-complete for all $k \geq 1$. The problems* SP-*(k) and* SC-*(k) are $\mathcal{PLS}$-complete for all $k \geq 2$. The problems* W3DM-*(k,l) and* X3C-*(k) are $\mathcal{PLS}$-complete for all $k \geq 6$ and $l \geq 12$.*

**Theorem 2.** *The problems* SP-*(1) and* SC-*(1) are polynomial-time solvable.*

All proofs can be found in Section 3. In Subsection 3.1, we investigate the $\mathcal{PLS}$-complexity of W3DM-$(p, q)$ and X3C-$(k)$, in Subsection 3.2, the $\mathcal{PLS}$-complexity of SP-$(k)$, in Subsection 3.3 the $\mathcal{PLS}$-complexity of SSP-$(k)$, in Subsection 3.4, the $\mathcal{PLS}$-complexity of SC-$(k)$, in Subsection 3.5 the $\mathcal{PLS}$-complexity of TS-$(k)$, in Subsection 3.6 the $\mathcal{PLS}$-complexity of SB-$(k)$, in Subsection 3.7 the $\mathcal{PLS}$-complexity of HS-$(k)$, in Subsection 3.8 the $\mathcal{PLS}$-complexity of IP-$(k)$, and in Subsection 3.9 the $\mathcal{PLS}$-complexity of CC-$(k)$, Let us remark that the reductions we present are *tight* in the sense of Schäffer and Yannakakis, [17].

*Neighborhoods, Weights, and Hardness.* The hardness of a PLS-problems crucially depends on both the structure of the neighborhood and the involved weights. On the one hand, if the neighborhood structure limits the options for improvements in every step such that this can be exploited by polynomial time algorithms, then the problems become easy, regardless of the weights. This is the case in SP-(1) and SC-(1) where the neighborhood structure can be exploited by a greedy algorithm. Interestingly, for all other problems we investigate, the neighborhood structure does not interfere with weights in terms of hardness. For most of the problems, this is the case even for the smallest possible neighborhood of size 1. On the other hand, if all weights are polynomially bounded then locally optimal solutions can be computed via successive improvements in polynomial time. All $\mathcal{PLS}$-complete generalized satisfiability problems we reduce from were proven to be $\mathcal{PLS}$-complete via tight reduction and the involved weights are of exponential size. We incorporate these weights in our reductions, preserving their overall structure. Usually, we introduce additional weights which are not part of the input problem. They belong to auxiliary gadgets that are specific to the reduction. The weights involved are either of size one or such that a single weight exceeds the sum of all weights in the original problem.

*The General Technique of Our Reductions.* As with most reductions in $\mathcal{PLS}$, [5, 17], our reductions for hardness results consist of two parts: In one part, we encode the input problem $I$ in the reduced instance $\Phi(I)$ in a rather direct manner, while preserving the structure of the original weights. In the other part, which is specific to the reduction and represents a large part of our contribution, we introduce auxiliary gadgets that enforce a particular structure in local optima. Eventually, these gadgets ensure that locally optimal solutions in $\Phi(I)$ indeed correspond to locally optimal solution in $I$. Our proofs also consist of two parts:

1. First, we show that all feasible solutions which are locally optimal for $\Phi(I)$ use the gadgets as intended, thereby uncovering the structure of locally optimal solutions. Depending on the reduction, we call these solutions *standard solutions* or to be *consistent* for some property.
2. Second, we show that all local optima for $\Phi(I)$ correspond to local optimal for $I$. Step 1 now allows to concentrate on the set of all consistent or standard solutions.

We want to stress that reducing from $(3, 2, r)$-MCA is crucial for us to show tight bounds for SETPACKING and SETCOVER. Furthermore, we believe that reducing from very restricted but $\mathcal{PLS}$-complete versions of the MAXCONSTRAINTASSIGNMENT-problem might prove useful for establishing that further $\mathcal{PLS}$-problems with a small neighborhood are $\mathcal{PLS}$-complete.

To the best of our knowledge, these are are one of the very few $\mathcal{PLS}$ results for local search on weighted standard set problems, as intensively studied in the literature. Our analysis also unveils that the hardness of the problems stems from the combination of a numerical problem on an underlying combinatorial problem.

## 3   The $\mathcal{PLS}$-Complexity of Weighted Standard Set Problems

In this section, we investigate the complexity of computing locally optimal solutions for the weighted standard set problems presented in Section 2.1.

***Preliminaries*** Denote by $(p,q,r)$-MINCA the minimization version of $(p,q,r)$-MCA. Here, results about $(p,q,r)$-MCA carry over to $(p,q,r)$-MINCA. In the following, let integer $W \in \mathbb{N}$ be larger than the sum of all weights in an instance $I$ from problem POSNAE, CNFSAT, $(p,q,r)$-MCA or $(p,q,r)$-MINCA. In detail, for a given instance of POSNAE or CNFSAT, let integer $W \gg \sum_{C_i \in \mathcal{C}} w_{C_i}$. For a given instance of $(p,q,r)$-MCA or $(p,q,r)$-MINCA, let integer $W \gg \sum_{C_i(x_{i_1},\dots x_{i_{p_i}}) \in \mathcal{C}} \sum_{\mathsf{a}(x_{i_1}),\dots,\mathsf{a}(x_{i_{p_i}}) \in [r]} w_{C_i}(\mathsf{a}(x_{i_1}),\dots,\mathsf{a}(x_{i_{p_i}}))$.

### 3.1   On the Complexity of W3DM-(p,q) and X3C-(k)

In this subsection, we show that W3DM-$(p,q)$ is $\mathcal{PLS}$-complete for all $p \geq 6$ and $q \geq 12$. Since instances of X3C-$(k)$ are instances of W3DM-$(k)$ by defining triples as 3-element sets, our reduction is also applicable to X3C-$(k)$ with the same argumentation. This eventually shows that X3C-$(k)$ is $\mathcal{PLS}$-complete for all $k \geq 6$. We present the reduction function $\Phi$ and the solution mapping $\Psi$, which are both slight modifications of a reduction proving that W3DM-$(9,15)$ is $\mathcal{PLS}$-complete, presented in [6]. We also use the notation presented therein.

***The Reduction*** In a nutshell, the main idea is to *mimic assignments of variables in a constraint with triples* possessing the weight of the constraint for the given assignment. An *additional gadget* ensures the *consistency* for all variable assignments.

In more detail, given an instance $I \in D_{(3,2,r)\text{-MCA}}$, we construct a reduced instance $\Phi(I) = (N,w) \in D_{\text{W3DM-}(6,12)}$, consisting of a positive integer $N \in \mathbb{N}$ and a weight function $w : [N]^3 \mapsto \mathbb{N}$ that maps triples to positive integer weights. From [5] it follows that the subclass of instances of $(3,2,r)$-MCA where every clause has length three and where the set of variables is tri-colored such that no clause contains two variables with the same color and all sets of variables with a certain color have the same cardinality is $\mathcal{PLS}$-complete. Thus, without loss of generality, we assume that in $I$, every constraint has length three, every variable appears twice and is colored blue, red, or white. The coloring of the variables is such that no clause contains two variables with the same color and each subset of variables with a certain color has cardinality $|\mathcal{X}|/3$. Let $\sigma$ be an ordering of $\mathcal{C}$. We define $N := 2 \cdot r \cdot |\mathcal{X}| + |\mathcal{X}|/3$ and introduce the following function $w$:

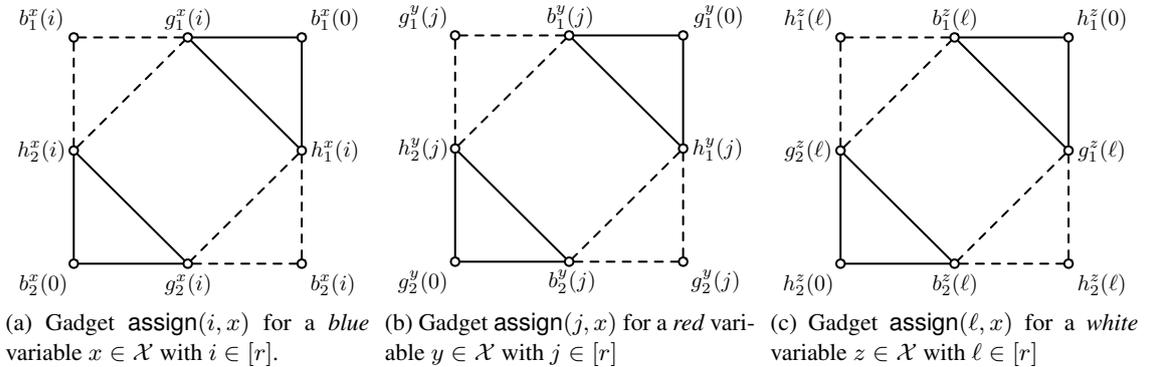

(a) Gadget $\mathsf{assign}(i,x)$ for a *blue* variable $x \in \mathcal{X}$ with $i \in [r]$.  (b) Gadget $\mathsf{assign}(j,x)$ for a *red* variable $y \in \mathcal{X}$ with $j \in [r]$.  (c) Gadget $\mathsf{assign}(\ell,x)$ for a *white* variable $z \in \mathcal{X}$ with $\ell \in [r]$

Fig. 1: Gadgets $\mathsf{assign}(i,x)$ for a blue, a red, and a white variable with two large triples (solid triangles) and two medium triples (dashed triangles).

*Forcing a consistent assignment* We define the three sets

$$B := \{b_s^x(i) \mid x \in X, i \in [r], s \in [2]\} \cup \{b_s^x(0) \mid x \in \mathcal{X} \text{ is a } \textit{blue} \text{ variable}, s \in [3]\},$$
$$G := \{g_s^x(i) \mid x \in X, i \in [r], s \in [2]\} \cup \{g_s^x(0) \mid x \in \mathcal{X} \text{ is a } \textit{red} \text{ variable}, s \in [3]\},$$
$$H := \{h_s^x(i) \mid x \in X, i \in [r], s \in [2]\} \cup \{h_s^x(0) \mid x \in \mathcal{X} \text{ is a } \textit{white} \text{ variable}, s \in [3]\},$$

each of cardinality $N$. For every *blue* variable $x \in \mathcal{X}$ and $i \in [r]$, we define gadgets assign$(i, x)$ consisting of two *large triples* $(b_1^x(0), g_1^x(i), h_1^x(i))$ and $(b_2^x(0), g_2^x(i), h_2^x(i))$ of weight $7W$ and two *medium triples* $(b_1^x(i), g_1^x(i), h_2^x(i))$ and $(b_2^x(i), g_2^x(i), h_1^x(i))$ of weight $3W$. We depicted a gadget assign$(i, x)$ in Figure 1a for some blue variable $x \in \mathcal{X}$ and $i \in [r]$. For every *red* variable $y \in \mathcal{X}$ and $j \in [r]$, we define gadgets assign$(j, y)$ consisting of two large triples $(b_1^y(j), g_1^y(0), h_1^y(j))$ and $(b_2^y(j), g_2^y(0), h_2^y(j))$ of weight $7W$ and two medium triples $(b_1^y(j), g_1^y(j), h_2^y(j))$ and $(b_2^y(j), g_2^y(j), h_1^y(j))$ of weight $3W$. We again depicted a gadget assign$(j, y)$ in Figure 1b for some red variable $y \in \mathcal{X}$ and $j \in [r]$. For every *white* variable $z \in \mathcal{X}$ and $\ell \in [r]$, we define gadgets assign$(\ell, z)$ consisting of two large triples $(b_1^z(\ell), g_1^z(\ell), h_1^z(0))$ and $(b_2^z(\ell), g_2^z(\ell), h_2^z(0))$ of weight $7W$ and two medium triples $(b_1^z(\ell), g_2^z(\ell), h_1^z(\ell))$ and $(b_2^z(\ell), g_1^z(\ell), h_2^z(\ell))$ of weight $3W$. We again depicted a gadget assign$(\ell, z)$ in Figure 1c for some white variable $z \in \mathcal{X}$ and $\ell \in [r]$.

*Evaluating the assignment* Without loss of generality, let $x \in \mathcal{X}$, be a blue variable, $y \in \mathcal{X}$ be a red variable, and $z \in \mathcal{X}$ be a white variable. For every constraint $C_i(x, y, z) \in \mathcal{C}$, where, with respect to $\sigma$, variable $x$ appears for the $s$-th, variable $y$ appears for the $t$-th time, and $z$ appears for the $u$-th time, with $s, t, u \in [2]$, we define *small* triples $(b_s^x(i), g_t^y(j), h_u^z(\ell))$ of weight $w_{C_i}(i, j, \ell)$ for every $i, j, \ell \in [r]$. All other triples have weight zero. This terminates the description of the reduction function $\Phi(I)$.

*Standard assignment* Extending the definition from [6], we define a *standard assignment* as a feasible solution $\mathsf{S} \in F_{\text{W3DM-}(6,12)}(\Phi(I))$, consisting of an *assignment part* and an *evaluation part*, of the following form: Considering the assignment part, for every blue variable $x \in \mathcal{X}$ there is some $i \in [r]$, such that for all $s \in [3]$, large triples $(b_s^x(0), g_s^x(i), h_s^x(i)) \in \mathsf{S}$. For all $j \in [r], j \neq i$, medium triples $(b_1^x(j), g_1^x(j), h_2^x(j)), (b_2^x(j), g_2^x(j), h_1^x(j)) \in \mathsf{S}$. Analogously, large and medium triples are present for red and white variables. Considering the evaluation part, let $x, y, z \in \mathcal{X}$ and $i, j, \ell \in [r]$, such that large triples for $x, y,$ and $z$ in $\mathsf{S}$ are from gadgets assign$(i, x)$, assign$(j, y)$, and assign$(\ell, z)$. For every constraint $C_p(x, y, z) \in \mathcal{C}$, where $x$ occurs for the $s$-th, $y$ occurs for the $t$-th time, and $z$ occurs for the $u$-th time, with respect to $\sigma$ and $s, t, u \in [3]$, if $x$ is a blue variable, $y$ is a red variable, and $z$ is a white variable, the triple $(b_s^x(i), g_t^y(j), h_u^z(\ell)) \in \mathsf{S}$; analogously for all other colorings of the involved variables.

*Solution Mapping* Again extending [6], if $\mathsf{S} \in F_{\text{W3DM-}(6,12)}(\Phi(I))$ is a standard assignment, then $\Psi(I, \mathsf{S})$ returns for every blue variable $x \in \mathcal{X}$ the index $i \in [r]$, such that $(b_1^x(0), g_1^x(i), h_1^x(i)) \in \mathsf{S}$, for every red variable $y \in \mathsf{X}$ the index $j \in [r]$, such that $(b_1^y(j), g_1^y(0), h_1^y(j)) \in \mathsf{S}$, and for every white variable $z \in \mathsf{X}$ the index $\ell \in [r]$, such that $(b_1^z(\ell), g_1^z(\ell), h_1^z(0)) \in \mathsf{S}$. If $\mathsf{S} \in F_{\text{W3DM-}(6,12)}(\Phi(I))$ is not a standard assignment, then $\Psi(I, \mathsf{S})$ returns the feasible solution computed by algorithm $\text{INIT}_{(3,2,r)\text{-MCA}}(I)$. This terminates the description of the reduction.

**Lemma 1.** *Every locally optimal solution* $\mathsf{S} \in F_{\text{W3DM-}(6,12)}(\Phi(I))$ *is a standard assignment.*

*Proof.* We present the proof for sake of completeness, as it is similar to the proof of Lemma 1 presented in [6]. Let $\mathsf{S} \in F_{\text{W3DM-}(6,12)}(\Phi(I))$ be a locally optimal solution. Without loss of generality, let $x \in \mathcal{X}$ be a blue variable.

***Roadmap*** With variable $x$ fixed, the proof splits into three parts:

1. We first show that there are two large triples $(b_1^x(0), g_1^x(i), h_1^x(i))$ and $(b_2^x(0), g_2^x(j), h_2^x(j))$ in S for some $i, j \in [r]$. For every gadget without a large triple, there are two medium triples in S.
2. Second, we prove that the two large triples are on the same gadget.
3. Finally, we show that the small triples in S are chosen in consistency with the placement of the large triples.

**(1)*: Two Large Triples and Two Medium Triples.*** Assume that w.l.o.g triple $(b_1^x(0), g_1^x(i), h_1^x(i)) \notin$ S for every $i \in [r]$. We construct a better solution that contains $(b_1^x(0), g_1^x(i), h_1^x(i))$ for some $i \in [r]$. On gadget $\mathsf{assign}(i, x)$, the large triple $(b_1^x(0), g_1^x(i), h_1^x(i))$ with weight $7W$ is built. The necessary elements $b_1^x(0)$, $g_1^x(i)$, and $h_1^x(i)$ are in at most three triples, each of weight at most $2W$. Thus, we substitute a total of three triples to obtain a strictly better solution. Considering the medium triples, assume that there exists some $j \in [r]$ such that no large triple and not both medium triples from gadget $\mathsf{assign}(j, x)$ are in S. Without loss of generality, let $(b_1^x(j), g_1^x(j), h_2^x(j)) \notin$ S. On gadget $\mathsf{assign}(j, x)$, the medium triple $(b_1^x(j), g_1^x(j), h_2^x(j))$ of weight $2W$ is built. The necessary elements are in at most three triples of total weight at most $W$. Thus, we again substitute a total of three triples to obtain a strictly better solution.

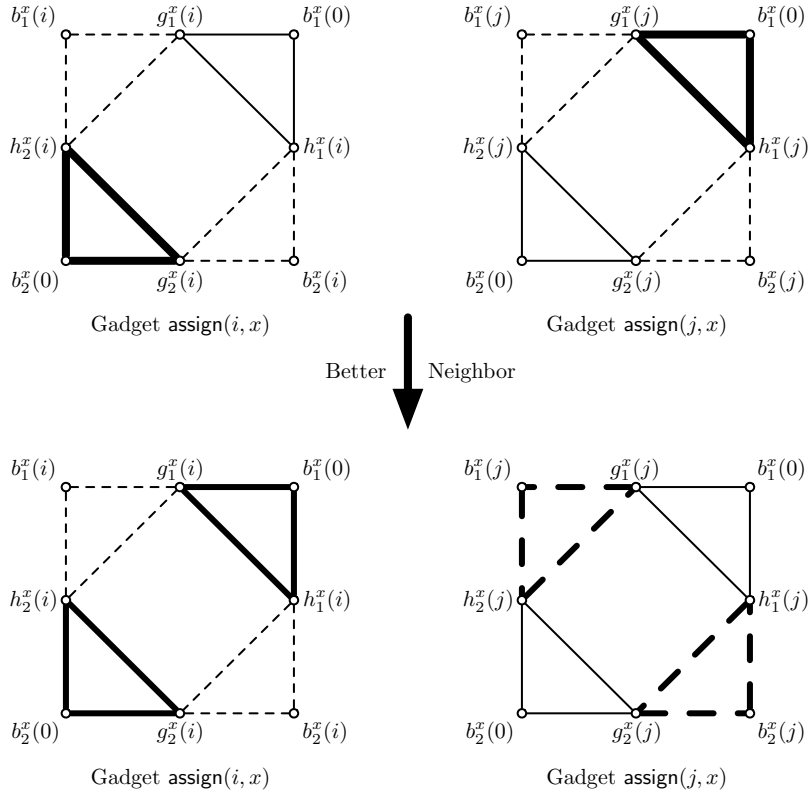

Fig. 2: Illustration of the construction of a better solution in (2) from proof of Lemma 1.

**(2)*: Two Large Triples On A Single Gadget.*** Assume that the large triples are placed on two different gadgets $\mathsf{assign}(i, x)$ and $\mathsf{assign}(j, x)$ for some $i, j \in [r]$ with $i \neq j$. In detail, let large triples $(b_2^x(0), g_2^x(i), h_2^x(i)) \in$ S and $(b_1^x(0), g_1^x(j), h_1^x(j)) \in$ S. We have depicted this situation in the upper part of Figure 2. Note that by construction, there are no medium triples from gadgets $\mathsf{assign}(i, x)$ or $\mathsf{assign}(j, x)$ in $\mathsf{S}$. We construct a better solution by removing the large triple $(b_1^x(0), g_1^x(j), h_1^x(j))$ from $\mathsf{S}$. Additionally, on gadget $\mathsf{assign}(i, x)$, the large triple $(b_1^x(0), g_1^x(i), h_1^x(i))$ with weight $7W$ is built. On gadget $\mathsf{assign}(j, x)$, the two new medium triples $(b_1^x(j), g_1^x(j), h_2^x(j))$ and $(b_2^x(j), g_2^x(j), h_1^x(j))$, each of weight $2W$, are built. We have depicted the better neighboring solution in the lower part of Figure 2. Elements $b_1^x(0), g_1^x(j)$, and $h_1^x(j)$) are in given triples. The remaining elements $g_1^x(i), h_1^x(i), b_1^x(j), h_2^x(j), b_2^x(j)$, and $g_2^x(j)$ are in at most six triples. Our construction yields an additional two medium triples, each of weight $2W$ while all decomposed triples which are not shifted to a different gadget have weight at most $W$. Thus, we replace a total of at most six triples to obtain a solution of strictly higher cost.

(3): *Small Weights.* The above two cases show that the assignment part of $\mathsf{S}$ is that of a standard assignment. In detail, for every variable $x \in \mathcal{X}$, there exists some $i \in [r]$ such that the two large triples are from the same gadget $\mathsf{assign}(i, x)$. For every blue variable $x$ this implies that elements $b_1^x(i)$ and $b_2^x(i)$ are not in any large or medium triple; analogously for every red and white variable. By construction, for every $C_i \in \mathcal{C}$, only one small triple with strictly positive weight can be uniquely chosen. Without loss of generality, let $x \in \mathcal{X}$ be a blue variable with $s \in [2]$ and $i \in [r]$ such that boy $b_s^x(i)$ is not in any large or medium triple. Without loss of generality, let $y \in \mathcal{X}$ be a red variable with $t \in [2]$ and $j \in [r]$ such that girl $g_t^y(j)$ is not in any large or medium triple. Without loss of generality, let $z \in \mathcal{X}$ be a white variable with $u \in [2]$ and $l \in [r]$ such that home $h_u^x(l)$ is not in any large or medium triple. Let $C_p(x, y, z) \in \mathcal{C}$ be such that $x$ appears for the $s$-th time, $y$ appears for the $t$-th time, and $z$ appears for the $u$-th time with respect to the given ordering $\sigma$. Assume that $\mathsf{S}$ deviates in the evaluation part. Thus, elements $b_s^x(i), g_t^y(j)$, and $h_u^x(\ell)$ are in at most three triples, each of weight zero. By building the small triple $(b_s^x(i), g_t^y(j), h_u^x(\ell))$ of weight $w_{C_p}(i, j, \ell)$, we replace at most three triples to obtain a neighboring solution with strictly improved cost. □

**Lemma 2.** $(3, 2, r)$-MCA $\leq_{pls}$ W3DM-$(p, q)$ *for all $p \geq 6$ and $q \geq 12$.*

*Proof.* Assume there exists a feasible solution $\mathsf{S} \in F_{\text{W3DM-}(6,12)}(\Phi(I))$ which is locally optimal for $\Phi(I)$, but is not locally optimal for $I$. By Lemma 1, $\mathsf{S}$ is a standard assignment. This implies that $\Psi(I, \mathsf{S})$ is a legal assignment to all variables $x \in \mathcal{X}$. Since $\Psi(I, \mathsf{S})$ is not locally optimal for $I$, there exists a (w.l.o.g.) white variable $z \in \mathcal{X}$ from instance $I \in (3, 2, r)$-MCA, which can be set from value $i \in [r]$ to a value $j \in [r]$ such that the objective function strictly increases by some $z > 0$. Let variable $z$ appear in constraints $C_p, C_q \in \mathcal{C}$. The neighboring solution of $\mathsf{S}$, where the two large triples are on gadget $\mathsf{assign}(j, z)$ and all medium triples are on gadgets $\mathsf{assign}(\ell, z)$ for all $\ell \in [r]$ and $\ell \neq j$, and all small triples are chosen according to the new assignment of value $j$ to $z$ improves the cost of $\mathsf{S}$ by $z$ by construction. This exchange involves the six triples $(*, *, h_1^z(0))$, $(*, *, h_2^z(0)), (*, *, h_1^z(i)), (*, *, h_2^z(i)), (*, *, h_1^z(j))$, and $(*, *, h_2^z(j))$. The involved homes are $h_1^z(i)$ and $h_2^z(i)$ from gadget $\mathsf{assign}(i, z)$, homes $h_1^z(j)$ and $h_2^z(j)$ from gadget $\mathsf{assign}(j, z)$ and homes $h_1^z(0)$ and $h_2^z(0)$ which are in every gadgets $\mathsf{assign}(*, z)$. On gadget $\mathsf{assign}(i, x)$, girl $g_2^z(i)$ and boy $b_1^z(i)$ move to home $h_1^z(i)$, and girl $g_1^z(i)$ and boy $b_2^z(i)$ move to home $h_2^z(i)$. On gadget $\mathsf{assign}(j, x)$, girl $g_1^z(j)$ and boy $b_1^z(j)$ move to home $h_1^z(0)$, and girl $g_2^z(j)$ and boy $b_2^z(j)$ move to home $h_2^z(0)$. All boys and girls in small triples move from homes $h_2^z(i)$ and $h_1^z(i)$ to respective homes $h_2^z(j)$ and $h_1^z(j)$. Thus, 12 boys or girls move to different homes. For all other colors of variables which switch assignment, at most 10 boys or girls move to new homes. □

### 3.2 The Exact Complexity of SETPACKING-$(k)$

In this subsection, we prove that SP-$(k)$ is $\mathcal{PLS}$-complete for all $k \geq 2$ and polynomial-time solvable for $k = 1$. Given an instance $I \in D_{(3,2,r)\text{-MCA}}$, we construct a reduced instance $\Phi(I) =$ $(\mathcal{M}, w) \in D_{\text{SP-}(2)}$, consisting of a collection $\mathcal{M}$ of sets over a finite set $\mathcal{B}$, a weight function $w : \mathcal{M} \mapsto \mathbb{N}$ that maps sets in collection $\mathcal{M}$ to positive integer weights, and a positive integer $m \leq |\mathcal{M}|$. W.l.o.g., we assume that in instance $I \in D_{(3,2,r)\text{-MCA}}$, every constraint $C_i \in \mathcal{C}$ has length 3 and the weight of every non-zero assignment is strictly larger than 1. Furthermore, we assume that every variable $x \in \mathcal{X}$ appears in 2 constraints and takes values from $[r]$. Additionally, we may assume that the variables are ordered by appearance.

***The Reduction*** In a nutshell, the main idea is to define *sets representing assignments of variables to values in constraints* such that *inconsistent assignments intersect*. The weight of a set corresponds to the weight of the constraint for the variable assignment the set represents. Additional *intersection-free sets* of weight 1 offer a relatively *small incentive* in situations where sets intersect.

In more detail, we create a reduced instance of SP-(2) with $m := |\mathcal{C}|$. Sets in collection $\mathcal{M}$ are defined on elements from the finite set $\mathcal{B} := \{e_i, c_i \mid i \in [m]\} \cup \{x_i \mid x \in \mathcal{X}, i \in [r]\}$. Collection $\mathcal{M}$ consists of the following sets: For all $i \in [m]$, we introduce sets $C_i^{SP} := \{e_i\}$ of weight $w(C_i^{SP}) := 1$ in $\mathcal{M}$. For every constraint $C_i(u, v, w) \in \mathcal{C}$ and every assignment $a, b, c \in [r]$, we introduce sets $C_i^{a,b,c}$ of weight $w(C_i^{a,b,c}) := w_{C_i}(a, b, c)$ in $\mathcal{M}$. Here, set $C_i^{a,b,c} := \{c_i, u'_a, v'_b, w'_c \mid a, b, c \in [r]\}$ and

$$u'_a := \begin{cases} u_a & \text{if } u \in \mathcal{X} \text{ appears in } C_i(u, v, w) \text{ for the first time} \\ u_1, \ldots, u_{a-1}, u_{a+1}, \ldots, u_r & \text{otherwise,} \end{cases}$$

analogously for $v'_b$ and $w'_c$. We call an element $x_j$ for some variable $x \in \mathcal{X}$ and asssignment $j \in [r]$ enclosed in a set from $\mathcal{M}$ due to the first appearance of $x$ *direct representative* of $x$. We say that a family of sets $C_i^{*,*,*} \in \mathcal{M}$ is *incident* to a family of sets $C_j^{*,*,*} \in \mathcal{M}$ if the clauses $C_i, C_j \in \mathcal{C}$ have a common variable.

***Solution Mapping*** We call a feasible solution $\mathsf{S} \in F_{\text{SP-}(2)}(\Phi(I))$ set-consistent if $|\mathsf{S}| = m$ and for every $i \in [m]$ there is exactly one set $C_i^{a,b,c}$ in $\mathsf{S}$ for some $a, b, c \in [r]$, which is pairwise disjoint from all other sets in $\mathsf{S}$. For a feasible and set-consistent solution $\mathsf{S}$, function $\Psi(I, \mathsf{S})$ returns for every set $C_i^{*,*,*} \in \mathsf{S}$ and every direct-representative of $x_j$ the assignment $j$ to variable $x \in \mathcal{X}$. If $\mathsf{S}$ is not set-consistent, then the assignment computed by $\text{INIT}_{(3,2,r)\text{-MCA}}(I)$ is returned.

**Lemma 3.** *Every locally optimal solution* $\mathsf{S} \in F_{\text{SP-}(2)}(\Phi(I))$ *is set-consistent.*

*Proof.* Assume there exists a locally optimal solution $\mathsf{S}' \in F_{\text{SP-}(2)}(\Phi(I))$ with $|\mathsf{S}'| < m$. By pigeonhole principle and construction of our reduction, this implies that there exists a set $C_j^{SP} \in \mathcal{M}$ with $j \in [m]$ which is not in $\mathsf{S}'$. Adding $C_j^{SP}$ to $\mathsf{S}'$ strictly improves the cost of $\mathsf{S}'$, since by construction $C_j^{SP}$ is pairwise disjoint from all sets in $\mathcal{M}$. A contradiction to $\mathsf{S}'$ being locally optimal.

Assume there exists a locally optimal solution $\mathsf{S}' \in F_{\text{SP-}(2)}(\Phi(I))$ with $|\mathsf{S}'| = m$ and there exists an $i \in [m]$ such that (1) at least two set $C_i^{a,b,c}, C_i^{d,e,f}$ are in $\mathsf{S}'$ for some $a, b, c, d, e, f \in [r]$ or (2) no set $C_i^{*,*,*}$ is in $\mathsf{S}'$. First, consider case (1). Note that sets $C_i^{a,b,c}, C_i^{d,e,f} \in \mathsf{S}'$ are not pairwise disjoint, since by construction they both contain element $c_i \in \mathcal{B}$. By definition of SP-$(k)$ this implies that they do not contribute to the cost of $\mathsf{S}'$. By pigeonhole principle and construction of our reduction there exists a set $C_j^{SP} \in \mathcal{M}$ for some $j \in [m]$ which is not in $\mathsf{S}'$. Replacing set $C_i^{d,e,f} \in \mathsf{S}'$ for set $C_j^{SP} \in \mathcal{M}$ strictly improves the cost of $\mathsf{S}'$, since $w(C_j^{SP}) = 1$ and $C_j^{SP}$ is pairwise disjoint from all sets in $\mathcal{M}$. Set $C_i^{a,b,c}$ may only become pairwise disjoint and thus contribute to the cost of $\mathsf{S}'$. A contradiction to $\mathsf{S}'$ being locally optimal. Now, consider case (2). Let $C_i^{*,*,*} \in \mathsf{S}'$ be incident to families of sets $C_o^{*,*,*}, C_p^{*,*,*}, C_q^{*,*,*} \in \mathcal{M}$. From case (1), we have that for every $j \in [m]$, at most one set $C_j^{*,*,*} \in \mathsf{S}'$. If present in $\mathsf{S}'$, assume that sets $C_o^{a,*,*}, C_p^{b,*,*}, C_q^{c,*,*}$ are in $\mathsf{S}'$ for some $a, b, c \in [r]$. Since there does not exist a set $C_i^{*,*,*} \in \mathsf{S}'$, this implies that there exists a set $C_j^{SP} \in \mathsf{S}'$ for some $j \in [m]$. Exchanging set $C_j^{SP} \in \mathsf{S}'$ for set $C_i^{a,b,c} \in \mathcal{M}$—if sets from incident families are not present in $\mathsf{S}'$, choose an arbirary value for the respective variable—strictly increases the cost of $\mathsf{S}'$, since $C_i^{a,b,c}$ is pairwise disjoint from all sets in $\mathsf{S}'$ and $w(C_i^{a,b,c}) > w(C_j^{SP})$. A contradiction to $\mathsf{S}'$ being locally optimal. □

**Lemma 4.** $(3, 2, r)$-MCA $\leq_{pls}$ SP-$(k)$ *for all* $r \in \mathbb{N}$, $k \geq 2$.

*Proof.* Assume there exists a feasible solution $\mathsf{S} \in F_{\text{SP-}(2)}(\Phi(I))$ which is locally optimal for $\Phi(I)$, but is not locally optimal for $I$. By Lemma 3, $\mathsf{S}$ is set-consistent. This implies that $\Psi(I, \mathsf{S})$ is a legal assignment of values to variables $x \in \mathcal{X}$. Since $\Psi(I, \mathsf{S})$ is not locally optimal for $I$, there exists a variable $x \in \mathcal{X}$ from instance $I \in (3, 2, r)$-MCA, which can be set from value $i \in [r]$ to some value $j \in [r]$ such that the objective function strictly increases by some $z > 0$. Let variable $x$ appear in constraints $C_p, C_q \in \mathcal{C}$. Exchanging the sets $C_p^{i,*,*}$ and $C_q^{i,*,*}$ by sets $C_p^{j,*,*}$ and $C_p^{j,*,*}$ in $\mathsf{S}$ yields a feasible and set-consistent solution and by construction this strictly increases the cost of $\mathsf{S}$ by $z$. A contradiction. □

Despite the negative result for SP-$(k)$ for all $k \geq 2$, it is possible to compute a locally optimal solution for all instances $I \in$ SP-$(1)$ in polynomial time.

**Lemma 5.** SP-$(1)$ *is polynomial-time solvable.*

*Proof.* Given an instance $I \in D_{\text{SP-}(1)}$, we use the following algorithm GREEDYPACKING: Starting from the feasible solution $\mathsf{S} := \emptyset$, process all sets in $\mathcal{M}$ by weight in descending order and add the heaviest yet unprocessed set to $\mathsf{S}$, if it is disjoint from all sets $S_i \in \mathsf{S}$. In order to prove that a solution $\mathsf{S} \in F_{\text{SP-}(1)}(I)$ computed by GREEDYPACKING is locally optimal, assume that GREEDYPACKING terminated and $\mathsf{S}$ is not locally optimal. This implies that there either exists a set $S_i \in \mathcal{M}$ that can be added, or a set $S_j \in \mathsf{S}$ that can be deleted, or exchanged for another set $S_\ell \in \mathcal{M}$ with $S_\ell \notin \mathsf{S}$. Assume there exists a set $S_i \in \mathcal{M}$ with $S_i \notin \mathsf{S}$ which can be added to $\mathsf{S}$ such that the cost strictly improves by some $z \in \mathbb{N}$. This implies that $S_i$ is pairwise disjoint from all sets from $\mathsf{S}$ and thus, GREEDYPACKING would have included set $S_i$. A contradiction. Assume there exists a set $S_j \in \mathsf{S}$ which can be deleted from $\mathsf{S}$ such that the cost strictly improves by some $z \in \mathbb{N}$. This implies that $S_j$ intersects with some set from $\mathsf{S}$ and GREEDYPACKING would have not included $S_j$. A contradiction. Assume there exists a set $S_j \in \mathsf{S}$ which can be exchanged for some set $S_\ell \in \mathcal{M}$ with $S_\ell \notin \mathsf{S}$ such that the cost strictly improves by some $z \in \mathbb{N}$. This implies that $S_\ell$ is pairwise disjoint from all sets in $\mathsf{S} \setminus S_j$ and has a larger weight than $S_j$. Thus, GREEDYPACKING would have included $S_\ell$ instead of $S_j$. A contradiction. □

### 3.3  On the Complexity of SETSPLITTING-$(k)$

In this subsection, we prove that SSP-$(k)$ is $\mathcal{PLS}$-complete for all $k \geq 1$. Given an instance $I \in$ POSNAE, we construct a reduced instance $\Phi(I) = (\mathcal{M}, w) \in D_{\text{SSP-}(1)}$ consisting of a collection $\mathcal{M}$ of sets over a finite set $\mathcal{B}$ and a weight function $w : \mathcal{M} \mapsto \mathbb{N}$ that maps sets from collection $\mathcal{M}$ to positive integer weights.

***The Reduction*** Since SETSPLITTING is similar to HYPERGRAPH-2-COLORABILITY, we use a direct reduction: From instance $I$, we define the reduced instance of SSP-$(1)$ over the finite set $\mathcal{B} := \mathcal{X}$. For every clause $C_i(x, y) \in \mathcal{C}$, we introduce sets $C_i^{SSp} := \{x, y\}$ in $\mathcal{M}$ and we define $w(C_i^{SSp}) := w_{C_i}$. Here, function $\Psi(I, \mathsf{S})$ returns for a feasible solution $\mathsf{S} \in F_{\text{SSP-}(1)}(\Phi(I))$ assignment 0 for all variables in $S_1$ and assignment 1 for all variables in $S_2$.

**Lemma 6.** POSNAE $\leq_{pls}$ SSP-$(k)$ *for all* $k \geq 1$.

*Proof.* Assume there exists a feasible solution $\mathsf{S} \in F_{\mathrm{SSP\text{-}(1)}}(\Phi(I))$ which is locally optimal for $\Phi(I)$, but is not locally optimal for $I$. This implies that there exists a variable $x \in \mathcal{X}$ in $I$ which can be flipped such that clauses $C_i, \ldots, C_j \in \mathcal{C}$ now have literals with non-identical assignments, clauses $C_p, \ldots, C_q \in \mathcal{C}$ now have literals with identical assignments and the cost of $\Psi(I, \mathsf{S})$ strictly increases by $z > 0$. By construction, this implies that in $\Phi(I)$, element $x \in \mathcal{B}$ can switch partition and now sets $C_i^{SSp}, \ldots, C_j^{SSp} \in \mathcal{M}$ are not entirely contained in either $S_1$ or $S_2$ and sets $C_p^{SSp}, \ldots, C_q^{SSp} \in \mathcal{M}$ are entirely contained in either $S_1$ or $S_2$. By definition of $w$, this strictly increases the cost of $\mathsf{S}$ by $z$. Thus, $\mathsf{S}$ is not locally optimal. A contradiction. □

### 3.4  The Exact Complexity of SETCOVER-$(k)$

In this subsection, we prove that SC-$(k)$ is $\mathcal{PLS}$-complete for all $k \geq 2$ and polynomial-time solvable for $k = 1$.

*The Reduction*  In a nutshell, the main idea is to *reuse the encoding of variable assignments and constraints* presented in Subsection 3.2 such that for every consistent assignment of variables to values, there exists a covering where no element is covered by two sets of the solution. Shifting the weights by a large constant incentivizes dropping sets which double cover elements.

In more detail, given an instance $I \in D_{(3,2,r)\text{-MINCA}}$, we construct a reduced instance $\Phi(I) = (\mathcal{M}, w) \in D_{\mathrm{SC\text{-}(2)}}$, consisting of a collection $\mathcal{M}$ of sets over a finite set $\mathcal{B}$, and a weight function $w : \mathcal{M} \mapsto \mathbb{N}$ that maps sets in collection $\mathcal{M}$ to positive integer weights. As in proof of Lemma 4, we assume that in instance $I$, every constraint $C_i \in \mathcal{C}$ has length 3, every variable $x \in \mathcal{X}$ appears in 2 constraints and takes values from $[r]$. Denote $m := |\mathcal{C}|$. We create a reduced instance of SC-$(2)$ over the finite set $\mathcal{B} := \{c_i \mid i \in [m]\} \cup \{x_i \mid x \in \mathcal{X}, i \in [r]\}$. For every constraint $C_i(u, v, w) \in \mathcal{C}$ and every assignment $a, b, c \in [r]$, we introduce sets $C_i^{a,b,c}$ of weight $w(C_i^{a,b,c}) := w_{C_i}(a, b, c) + W$ in $\mathcal{M}$. Here, sets $C_i^{a,b,c}$ are defined as in proof of Lemma 4. The definition of an incident familiy, a set-consistent solution, and the solution mapping $\Psi(I, \mathsf{S})$ is as in proof of Lemma 4, except that for a non-set-consistent solution, now the assignment computed by $\text{INIT}_{(3,2,r)\text{-MINCA}}(I)$ is returned.

**Lemma 7.** *Every locally optimal solution* $\mathsf{S} \in F_{\mathrm{SC\text{-}(2)}}(\Phi(I))$ *is set-consistent.*

*Proof.* Assume there exists a locally optimal solution $\mathsf{S}' \in F_{\mathrm{SC\text{-}(2)}}(\Phi(I))$ with $|\mathsf{S}'| < m$. By pigeonhole principle and construction of our reduction, this implies that there exists an element $c_i \in \mathcal{B}$ for some $i \in [m]$ which is not covered. A contradiction to $\mathsf{S}'$ being a feasible solution.

Assume there exists a locally optimal solution $\mathsf{S}' \in F_{\mathrm{SC\text{-}(2)}}(\Phi(I))$ with $|\mathsf{S}'| > m$. By pigeonhole principle, this implies that there are two sets $C_i^{a,b,c}$ and $C_i^{d,e,f}$ with total weight at least $2W$ in $\mathsf{S}$ for some $a, b, c, d, e, f \in [r]$ and $i \in [m]$. Since $\mathsf{S}'$ is a feasible solution, there exist sets $C_h^{o,*,*}, C_j^{p,*,*}, C_l^{q,*,*} \in \mathsf{S}'$ for some $o, p, q \in [r]$ and $h, j, l \in [m]$ from families indicent to sets $C_i^{a,b,c}, C_i^{d,e,f} \in \mathsf{S}'$. Exchanging the two sets $C_i^{a,b,c}, C_i^{d,e,f}$ for set $C_i^{o,p,q}$ yields a feasible solution and strictly decreases the cost of $\mathsf{S}'$, since $w(C_i^{o,p,q}) < 2W$. A contradiction. □

**Lemma 8.** $(3, 2, r)$-MINCA $\leq_{pls}$ SC-$(k)$ *for all* $r \in \mathbb{N}$, $k \geq 2$.

*Proof.* Assume there exists a feasible solution $\mathsf{S} \in F_{\mathrm{SC\text{-}(2)}}(\Phi(I))$ which is locally optimal for $\Phi(I)$, but is not locally optimal for $I$. By Lemma 7, $\mathsf{S}$ is a set-consistent assignment. This implies that $\Psi(I, \mathsf{S})$ is a legal assignment of values to variables $x \in \mathcal{X}$. Since $\Psi(I, \mathsf{S})$ is not locally optimal for $I$, there exists a variable $x \in \mathcal{X}$ from instance $I \in (3, 2, r)$-MINCA, which can be set from value $i \in [r]$ to a value $j \in [r]$ such that the objective function strictly increases by some $z > 0$. Let variable $x$ appear in constraints $C_p, C_q \in \mathcal{C}$. Exchanging sets $C_p^{i,*,*}$ and $C_q^{i,*,*}$ for sets $C_p^{j,*,*}$ and $C_q^{j,*,*}$ in S yields a feasible and set-consistent solution and this strictly decreases the cost of S by $z$, by construction. A contradiction. □

Despite the negative result for SC-$(k)$ for all $k \geq 2$, it is again possible to compute a locally optimal solution for all instances $I \in \text{SP-}(1)$ in polynomial time.

**Lemma 9.** *SC-$(1)$ is polynomial-time solvable.*

*Proof.* Given an instance $I \in D_{\text{SC-}(1)}$, we use the following algorithm GREEDYCOVER: Starting from the initial feasible solution $\mathsf{S} := \mathcal{M}$, process all sets in S by weight in descending order and remove the heaviest yet unprocessed set if S is still a legal cover of $\mathcal{B}$ after the removal. In order to prove that a solution $\mathsf{S} \in F_{\text{SP-}(1)}(I)$ computed by GREEDYCOVER is locally optimal, assume that GREEDYCOVER terminated and S is not locally optimal. This implies that there exists a set $S_i \in \mathsf{S}$ that can be deleted or exchanged for another set $S_j \in \mathcal{M}$ with $S_j \notin \mathsf{S}$ such that the cost strictly improves by some $z > 0$. Assume there exists a set $S_i \in \mathsf{S}$ which can be removed. This implies that S is still a legal cover of $\mathcal{B}$ after the removal of $S_i$ and thus, GREEDYCOVER would have removed set $S_i$ as well. A contradiction. Assume there exists a set $S_i \in \mathsf{S}$ which can be exchanged for a set $S_j \in \mathcal{M}$ with $S_j \notin \mathsf{S}$. This implies that set $S_j$ of smaller weight covers all elements $\mathcal{B} \setminus \bigcup_{S_\ell \in (\mathsf{S} \setminus S_i)} S_\ell$, i.e. all elements which are uncovered if $S_i$ would be removed from S. Since $S_i$ has larger weight and after its removal, S is still a legal cover, GREEDYCOVER would have deleted $S_i$ from S. A contradiction. □

### 3.5 On the Complexity of TESTSET-$(k)$

In this subsection, we prove that TS-$(k)$ is $\mathcal{PLS}$-complete for all $k \geq 1$. Given an instance $I \in \text{POSNAE}$, we construct an instance $\Phi(I) = (\mathcal{M}, w, m) \in D_{\text{TS-}(1)}$. Here, $\Phi(I)$ consists of a collection $\mathcal{M}$ of sets over a finite set $\mathcal{B}$, a weight function $w : \mathcal{M} \times \mathcal{M} \mapsto \mathbb{N}$ that maps tuples of elements of $\mathcal{B}$ to positive integer weights, and a positive integer $m \leq |\mathcal{M}|$.

*The Reduction* The main idea is on the one hand to *encode the assignment of variables to values in the choice of singleton sets representing literals* in some solution and on the other hand to *simulate the evaluation of clauses in the weight function $w$*. Additional *small incentives* reward the *inclusion of singleton sets* whereas *medium incentives* reward the *inclusion of distinct literals for variables*.

In more detail, given instance $I$, we construct the reduced instance of TS-$(1)$ over the finite set $\mathcal{B} := \{x_0, x_1 \mid x \in \mathcal{X}\}$. We set $m := |\mathcal{X}|$ and define $\mathcal{M} := \{\{x_0\}, \{x_1\} \mid x \in \mathcal{X}\}$. For every clause $C_i(x, y) \in \mathcal{C}$, we define $w(x_0, y_1) = w(x_1, y_0) := w_{C_i} + 1 + W$. For all pairs $(x_i, y_i)$ with $x_i, y_i \in \mathcal{B}$, $i \in \{0, 1\}$, $x_i \neq y_{\bar{i}}$, and $x_i \neq y_i$ where there does not exist a clause $C_j(x, y) \in \mathcal{C}$, we define $w(x, y) := W + 1$. All other function values of $w$ are defined as 1. We call a feasible solution S *positive-element-consistent* if $|\mathsf{S}| = m$ and for every set $\{x_i\} \in \mathsf{S}$ it holds that $\{x_{\bar{i}}\} \notin \mathsf{S}$. Here, function $\Psi(I, \mathsf{S})$ returns for every feasible and positive-element-consistent solution $\mathsf{S} \in F_{\text{TS-}(1)}(\Phi(I))$ the assignment induced by the indices of the elements in the sets in S. If S is not positive-element-consistent, the assignment computed by $\text{INIT}_{\text{POSNAE}}(I)$ is returned.

**Lemma 10.** *Every locally optimal solution $\mathsf{S} \in F_{\text{TS-}(1)}(\Phi(I))$ is positive-element-consistent.*

*Proof.* Assume there exists a locally optimal solution $\mathsf{S}' \in F_{\text{TS-}(1)}(\Phi(I))$ with $|\mathsf{S}'| < m$. Since by construction $|\mathcal{M}| = 2m$, there exists a sets $\{x\} \in \mathcal{M}$ with $\{x\} \notin \mathsf{S}'$. Adding $\{x\}$ to $\mathsf{S}'$ increases the cost of $\mathsf{S}'$ by at least 1, since $w(x, r) \geq 1$ for all $r \in \mathcal{B}$. A contradiction.

Assume there exists a locally optimal solution $\mathsf{S}' \in F_{\text{TS-(1)}}(\Phi(I))$ with $|\mathsf{S}'| = m$, $\mathsf{S}'$ contains two sets $\{x_0\} \in \mathsf{S}'$ and $\{x_1\} \in \mathsf{S}'$. By pigeonhole principle, there exist sets $\{y_0\} \in \mathcal{M}$ and $\{y_1\} \in \mathcal{M}$ with $\{y_0\} \notin \mathsf{S}'$ and $\{y_1\} \notin \mathsf{S}'$. Thus, exchanging $\{x_0\}$ for $\{y_0\}$ increases the cost of $\mathsf{S}'$, since no weight $W$ is lost and the additional weight of $w(x_1, y_0) \geq W$ dominates the sum of the weights lost due to the removal of $\{x_0\}$. A contradiction. □

**Lemma 11.** POSNAE $\leq_{pls}$ TS-$(k)$ *for all* $k \geq 1$.

*Proof.* Assume there exists a feasible solution $\mathsf{S} \in F_{\text{TS-(1)}}(\Phi(I))$ which is locally optimal for $\Phi(I)$, but is not locally optimal for $I$. By Lemma 10, $\mathsf{S}$ is positive-element-consistent. This implies that $\Psi(I, \mathsf{S})$ is a legal assignment for variables $x \in \mathcal{X}$. Since $\Psi(I, \mathsf{S})$ is not locally optimal for $I$, there exists a variable $x \in \mathcal{X}$ from instance $I$, which can be flipped such that clauses $C_i, \ldots, C_j \in \mathcal{C}$ now have literals with non-identical assignments, clauses $C_p, \ldots, C_q \in \mathcal{C}$ now have literals with identical assignments and the cost strictly increases by $z > 0$. This implies that in $\Phi(I)$, set $\{x_i\} \in \mathsf{S}$ can be replaced by set $\{x_{\bar{i}}\} \in \mathcal{M}$. On the one hand, for every variable $y \in \mathcal{X}$ with $y \neq x$ which appears in a clause $C_l$ from $\{C_i, \ldots, C_j\}$ we have that element $y_i \in \mathsf{S}$ and $w(x_{\bar{i}}, y_i) := w_{C_l} + 1 + W$. On the other hand, for every variable $z \in \mathcal{X}$ with $z \neq x$ which appears in a clause $C_t$ from $\{C_p, \ldots, C_q\}$ we have that element $y_{\bar{i}} \in \mathsf{S}$ and $w(x_{\bar{i}}, y_{\bar{i}}) := W + 1$ by construction. All other pairs of elements of $\mathcal{B}$ remain unchanged in $\mathsf{S}$. By definition of $w$, this strictly increases the cost of $\mathsf{S}$ by $z$. Thus, $\mathsf{S}$ is not locally optimal. A contradiction. □

### 3.6  On the Complexity of SETBASIS-$(k)$

In this subsection, we prove that SB-$(k)$ is $\mathcal{PLS}$-complete for all $k \geq 1$. Given an instance $I \in (h)$-CNFSAT, we construct an instance $\Phi(I) = (\mathcal{M}, w, m) \in D_{\text{SB-(1)}}$ consisting of a collection $\mathcal{M}$ of sets over a finite set $\mathcal{B}$, a weight function $w : \mathcal{M} \mapsto \mathbb{N}$ that maps sets in collection $\mathcal{M}$ to positive integer weights, and a positive integer $m \leq |\mathcal{M}|$.

*The Reduction* In a nutshell, the main idea is to *encode every satisfying assignment of a clause via sets containing the respective literals and possessing the weight of the clause*. This is polynomial in the size of the input since the length of a clause in $I$ is at most $h$. For feasible solutions to be a collection of singleton sets, we add *large incentives* to include *singleton sets* and *medium incentives* to include *literals for distinct variables*.

From instance $I$, we construct a reduced instance of SB-(1) over the finite set $\mathcal{B} := \{x, \bar{x} \mid x \in \mathcal{X}\}$ and we define $m := |\mathcal{X}|$. For every $x \in \mathcal{X}$, we introduce sets $C_x^{SB} := \{x\}$, $C_{\bar{x}}^{SB} := \{\bar{x}\}$ in $\mathcal{M}$ with weight $w(C_x^{SB}) = w(C_{\bar{x}}^{SB}) := 2W$. For every $x, y \in \mathcal{X}$ with $x \neq y$ and $y \neq \bar{x}$, we introduce sets $C_{xy}^{SB} := \{x, y\}$, $C_{\bar{x}y}^{SB} := \{\bar{x}, y\}$, $C_{x\bar{y}}^{SB} := \{x, \bar{y}\}$, and $C_{\bar{x}\bar{y}}^{SB} := \{\bar{x}, \bar{y}\}$ in $\mathcal{M}$ which all have weight $w(\cdot) = W$. We encode every satisfying assignment for every clause $C_i \in \mathcal{C}$ by the respective set, define it to possess the weight of clause $C_i$, and add it to $\mathcal{M}$. In detail, for every clause $C_i(x_{i_1}, \ldots, x_{i_k}) \in \mathcal{C}$ and every satisfying assignment $\varphi : \{x_{i_1}, \ldots, x_{i_k}\} \mapsto \{0, 1\}^{i_k}$ for $C_i$, we introduce sets $C_i^{SB[\varphi(x_{i_1}, \ldots, x_{i_k})]}$ in $\mathcal{M}$ with weight $w(C_i^{SB[\varphi(x_{i_1}, \ldots, x_{i_k})]}) := w_{C_i}$. Here, $C_i^{SB[\varphi(x_{i_1}, \ldots, x_{i_k})]} := \{\varphi'(x_{i_1}), \ldots, \varphi'(x_{i_j}), \ldots, \varphi'(x_{i_k})\}$, where $\varphi'(x_i) := x_i$ if $P_i(\varphi(x_{i_1}, \ldots, x_{i_k})) = 1$ and $\varphi'(x_i) := \bar{x}_i$ otherwise. We call a feasible solution $\mathsf{S}$ *single-set-consistent* if $|\mathsf{S}| = m$, $|S_i| = 1$ for all $S_i \in \mathsf{S}$ and for every set $\{x_i\} \in \mathsf{S}$ it holds that $\{\bar{x}_i\} \notin \mathsf{S}$. Here, function $\Psi(I, \mathsf{S})$ returns for a feasible and element-consistent solution $\mathsf{S} \in F_{\text{SB-(1)}}(\Phi(I))$ for every set $\{x\} \in \mathsf{S}$ assignment 1 for variable $x \in \mathcal{X}$ and for every set $\{\bar{x}\} \in \mathsf{S}$ assignment 0 for variable $x \in \mathcal{X}$. If $\mathsf{S}$ is not single-set-consistent, the assignment computed by $\text{INIT}_{(h)\text{-CNFSAT}}(I)$ is returned.

**Lemma 12.** *Every locally optimal solution* $\mathsf{S} \in F_{\text{SB-(1)}}(\Phi(I))$ *is single-set-consistent.*

*Proof.* Assume there exists a locally optimal solution $\mathsf{S}' \in F_{\text{SB-}(1)}(\Phi(I))$ which contains a set $S_i \in \mathsf{S}'$ with $|S_i| > 1$. Since set $S_i$ has cardinality at least two, $S_i$ is only used in the union of sets which by construction have total weight strictly smaller than $2W$. By pigeonhole principle, there exists a set $\{x\} \in 2^{|\mathcal{B}|}$ which is not in $\mathsf{S}'$. Exchanging set $S_i$ for $\{x\}$ strictly increases the cost function. Now, set $C_x^{SB}$ can be constructed and $w(C_x^{SB})$ is larger than the sum of the weights of the sets which cannot be generated any more due to the removal of $S_i$. Thus, $\mathsf{S}'$ is not locally optimal. A contradiction.

Assume there exists a locally optimal solution $\mathsf{S}' \in F_{\text{SB-}(1)}(\Phi(I))$ and $\mathsf{S}'$ contains two sets $\{x\}, \{\bar{x}\} \in \mathsf{S}'$. By pigeonhole principle, there exists a set $\{y\} \in \mathcal{B}$ with $\{y\} \notin \mathsf{S}'$ and $\{\bar{y}\} \notin \mathsf{S}'$. Thus, by exchanging $\{\bar{x}\}$ for $\{y\}$, the additional set $C_{xy}^{SB}$ can now be constructed. No weight $2W$ is lost due to the exchange operation. Set $C_{xy}^{SB}$ has weight $W$ and this dominates the sum of the weights of the sets which may not be constructed any more due to the removal of $\{\bar{x}\}$. Thus, the cost of $\mathsf{S}'$ increased and $\mathsf{S}'$ is not locally optimal. A contradiction. □

**Lemma 13.** $(h)$-CNFSAT $\leq_{pls}$ SB-$(k)$ *for all* $k \geq 1$.

*Proof.* Assume there exists a feasible solution $\mathsf{S} \in F_{\text{SB-}(1)}(\Phi(I))$ which is locally optimal for $\Phi(I)$, but is not locally optimal for $I$. By Lemma 12, $\mathsf{S}$ is single-set-consistent. This implies that $\Psi(I, \mathsf{S})$ is a legal assignment for all variables $x \in \mathcal{X}$. Since $\Psi(I, \mathsf{S})$ is not locally optimal for $I$, there exists a variable $x \in \mathcal{X}$ in (h)-CNFSAT, which can be flipped such that clauses $C_i, \ldots, C_j \in \mathcal{C}$ become satisfied, clauses $C_p, \ldots, C_q \in \mathcal{C}$ become unsatisfied and the cost strictly increases by $z > 0$. This implies that in $\Phi(I)$, set $\{x\} \in 2^{|\mathcal{B}|}$ can be replaced by set $\{\bar{x}\} \in 2^{|\mathcal{B}|}$. Now, sets $C_i^{SB[\varphi(x_{i_1},\ldots,x_{i_k})]}, \ldots, C_j^{SB[\varphi(x_{j_1},\ldots,x_{j_k})]} \in \mathcal{M}$ corresponding to the satisfying assignment for the respective clauses can be generated by the union of subcollections of sets of $\mathsf{S}$ involving $\{\bar{x}\}$, and all sets $C_p^{SB[\varphi(x_{i_1},\ldots,x_{p_k})]}, \ldots, C_q^{SB[\varphi(x_{q_1},\ldots,x_{q_k})]} \in \mathcal{M}$ cannot be formed by the union of subcollections of sets of $\mathsf{S}$. By definition of $w$, this strictly increases the cost of $\mathsf{S}$ by $z$. Thus, $\mathsf{S}$ is not locally optimal. A contradiction. □

### 3.7 On the Complexity of HITTINGSET-($k$)

In this subsection, we prove that HS-$(k)$ is $\mathcal{PLS}$-complete for all $k \geq 1$. Given an instance $I \in$ CNFSAT, we construct an instance $\Phi(I) = (\mathcal{M}, w, m) \in D_{\text{HS-}(1)}$ consisting of a collection $\mathcal{M}$ of sets over a finite set $\mathcal{B}$, a weight function $w : \mathcal{M} \mapsto \mathbb{N}$ mapping sets in collection $\mathcal{M}$ to positive integer weights, and a positive integer $m \leq |\mathcal{B}|$.

***The Reduction*** In a nutshell, the main idea is to *encode every clause as some set containing the respective literals and possessing the weight of the clause*. To ensure consistency, we add *large incentives* to *include at least one literal* from every variable, *but not both*.

In more detail, from instance $I$, we create a reduced instance of HS-$(1)$ over the finite set $\mathcal{B} := \{x, \bar{x} \mid x \in \mathcal{X}\}$ where we define $m := |\mathcal{X}|$. For every variable $x \in \mathcal{X}$, we introduce sets $C_x^{HS} := \{x, \bar{x}\}$ in $\mathcal{M}$ with $w(C_x^{HS}) := W$. For every clause, we introduce a single set possessing the weight of the respective clause and the elements of the set correspond to the literals in the clause. In detail, for every clause $C_i(x_{i_1}, \ldots, x_{i_l}) \in \mathcal{C}$, we introduce sets $C_i^{HS} := \{x_{i_1}, \ldots, x_{i_l}\}$ in $\mathcal{M}$, and we define $w(C_i^{HS}) := w_{C_i}$. We call a feasible solution $\mathsf{S}$ *element-consistent* if $|\mathsf{S}| = |\mathcal{C}|$ and for every element $x \in \mathsf{S}$ it holds that $\bar{x} \notin \mathsf{S}$. Here, function $\Psi(I, \mathsf{S})$ returns for a feasible and element-consistent solution $\mathsf{S} \in F_{\text{HS-}(1)}(\Phi(I))$ for every element $x \in \mathsf{S}$ assignment 1 for variable $x \in \mathcal{X}$ and for every element $\bar{x} \in \mathsf{S}$ assignment 0 for variable $x \in \mathcal{X}$. If $\mathsf{S}$ is not element-consistent, then the assignment computed by INIT$_{\text{CNFSAT}}(I)$ is returned.

**Lemma 14.** *For every locally optimal solution,* $\mathsf{S} \in F_{\text{HS-}(1)}(\Phi(I))$ *is element-consistent.*

*Proof.* Assume there exists a locally optimal solution $\mathsf{S}' \in F_{\mathrm{HS}\text{-}(1)}(\Phi(I))$ with $|\mathsf{S}'| < m$. Since $|\mathcal{B}| = 2m$, there exists an element $x \in \mathcal{B}$ with $x \notin \mathsf{S}'$ and $\bar{x} \notin \mathsf{S}'$. Thus, adding $x$ to $\mathsf{S}'$ increases the cost of $\mathsf{S}'$, since no weight is lost and set $w(C_x^{HS})$ is hit. A contradiction.

Assume there exists a locally optimal solution $\mathsf{S}' \in F_{\mathrm{HS}\text{-}(1)}(\Phi(I))$ with $|\mathsf{S}'| = m$ and $\mathsf{S}'$ contains two elements $x \in \mathsf{S}'$ and $\bar{x} \in \mathsf{S}'$. By pigeonhole principle, there exists an element $y \in \mathcal{B}$ with $y \notin \mathsf{S}'$. Exchanging $x$ for $y$ in solution $\mathsf{S}'$ increases the cost of $\mathsf{S}'$. All sets of weight $W$ that were previously hit are still hit. Additionally, set $C_y^{HS}$ is now hit and $w(C_y^{HS})$ is larger than the sum of all sets that where hit due to the membership of $x \in \mathsf{S}'$. A contradiction. □

**Lemma 15.** CNFSAT $\leq_{pls}$ HS-$(k)$ *for all $k \geq 1$.*

*Proof.* Assume there exists a feasible solution $\mathsf{S} \in F_{\mathrm{HS}\text{-}(1)}(\Phi(I))$ which is locally optimal for $\Phi(I)$, but is not locally optimal for $I$. By Lemma 14, $\mathsf{S}$ is element-consistent. This implies that $\Psi(I, \mathsf{S})$ is a legal assignment to all variables $x \in \mathcal{X}$. Since $\Psi(I, \mathsf{S})$ is not locally optimal for $I$, there exists a variable $x \in \mathcal{X}$ in CNFSAT, which can be flipped such that clauses $C_i, \ldots, C_j \in \mathcal{C}$ become satisfied, clauses $C_p, \ldots, C_q \in \mathcal{C}$ become unsatisfied and the cost strictly increases by $z > 0$. This implies that in $\Phi(I)$, element $x \in \mathsf{S}$ can be replaced by element $\bar{x} \in \mathcal{B}$ and now sets $C_i^{HS}, \ldots, C_j^{HS} \in \mathcal{M}$ are hit, and sets $C_p^{HS}, \ldots, C_q^{HS} \in \mathcal{M}$ are not hit. By definition of $w$, this strictly increases the cost of $\mathsf{S}$ by $z$. Thus, $\mathsf{S}$ is not locally optimal. A contradiction. □

### 3.8  On the Complexity of INTERSECTIONPATTERN-$(k)$

In this subsection, we prove that IP-$(k)$ is $\mathcal{PLS}$-complete for all $k \geq 1$. Given an instance $I \in$ POSNAE, we construct an instance $\Phi(I) = (A, B, \mathcal{M}) \in D_{\mathrm{IP}\text{-}(1)}$ consisting of two symmetric $n \times n$ matrices $A := (a_{ij})_{i,j \in [n]}$, and $B := (b_{ij})_{i,j \in [n]}$, and a collection $\mathcal{M}$ of sets over a finite set $\mathcal{B}$. W.l.o.g., we assume that in $I$, every pair of variables $x, y \in \mathcal{X}$ occurs in some clause $C_i(x, y) \in \mathcal{C}$. Furthermore, let $\sigma$ be an ordering of $\mathcal{X}$ and let $\gamma_{x_i}$ denote the number of clauses $C_j \in \mathcal{C}$ variable $x_i \in \mathcal{X}$ appears in.

*The Reduction* In a nutshell, the main idea is for every variable $x \in \mathcal{X}$ to introduce *sets of identical cardinality* for *both assignments*, but which have *distinct* cardinality from all other sets. These sets contain *elements encoding satisfying assignments for all clauses variable $x$ appears in*. If a clause is satisfied by a given assignment, then the intersection of the two corresponding sets has cardinality two. In this case, the weight of the clause is added to the solution. Large incentives ensure that, identified by cardinality, the sets for variables are placed in the right position in the solution.

In more detail, let $v := |\mathcal{X}|$ and $m := 2|\mathcal{C}|$. We create a reduced instance of IP-$(1)$ over the finite set $\mathcal{B} := \{x_i^{C_j} \mid x \in \mathcal{X}, i \in \{0,1\}, x \text{ appears in clause } C_j \in \mathcal{C}\} \cup \{x_i^l \mid x \in \mathcal{X}, i \in \{0,1\}, l \in [m - 2\gamma_{x_i} + \sigma(x)]\}$. For every variable $x \in \mathcal{X}$ and $i \in \{0,1\}$, we introduce sets $C_{x_i}^{IP}$ in $\mathcal{M}$, where $C_{x_i}^{IP} := \{x_i^{C_j}, y_{\bar{i}}^{C_j} \mid x \text{ appears in clause } C_j(x, y) \in \mathcal{C}\} \cup \{x_i^l \mid l \in [m - 2\gamma_{x_i} + \sigma(x)]\}$. Note that by construction, for every $x \in \mathcal{X}$, $|C_{x_*}^{IP}| = m + \sigma(x)$. In the $v \times v$ matrix $A$ for $\Phi(I)$, we define $a_{ii} := m + i$ for all $i \in [n]$ and for all $i, j \in [n]$ with $i \neq j$, we define $a_{ij} := 2$. In the $v \times v$ matrix $B$ for $\Phi(I)$, we define $b_{ii} := W$ for all $i \in [n]$ and for all $i, j \in [n]$ with $i \neq j$, we define $b_{ij} := w_{C_k}$, where $w_{C_k}$ is the weight of clause $C_k(x_i, x_j) \in \mathcal{C}$. We say that a feasible solution $\mathsf{S} \in F_{\mathrm{IP}\text{-}(1)}(\Phi(I))$ is *position-consistent* if for every $i \in [n]$, set $C_{x_*}^{IP}$ on position $i$ has cardinality $a_{ii}$. Here, function $\Psi(I, \mathsf{S})$ returns for a feasible and position-consistent solution $\mathsf{S} \in F_{\mathrm{IP}\text{-}(1)}(\Phi(I))$ for every $C_{x_i}^{IP} \in \mathsf{S}$ assignment $i$ for variable $x \in \mathcal{X}$. If $\mathsf{S}$ is not position-consistent, then the assignment computed by $\mathrm{INIT}_{\mathrm{POSNAE}}(I)$ is returned.

**Lemma 16.** *Every locally optimal solution $\mathsf{S} \in F_{\mathrm{IP}\text{-}(1)}(\Phi(I))$ is position-consistent.*

*Proof.* Assume there exists a locally optimal solution $\mathsf{S}' \in F_{\text{IP-(1)}}(\Phi(I))$ which is not position-consistent. This implies that there exists a set $C_{x_*}^{IP} \in \mathsf{S}'$ on position $i \in [v]$ with $|C_{x_*}^{IP}| \neq a_{ii}$. Replacing $C_{x_*}^{IP} \in \mathsf{S}'$ for set $C_{y_*}^{IP} \in \mathcal{M}$ on position $i$ in $\mathsf{S}$ with $|C_{y_*}^{IP}| = a_{ii}$ strictly improves the cost of $\mathsf{S}'$, since $b_{ii} > \sum_{p<q\in[n], p\neq q} b_{pq}$ and entries $b_{pq}$ are the only terms that may be lost in the cost of $\mathsf{S}'$ due to the replacement. A contradiction. □

**Lemma 17.** POSNAE $\leq_{pls}$ IP-$(k)$ *for all* $k \geq 1$.

*Proof.* Assume there exists a feasible solution $\mathsf{S} \in F_{\text{IP-(1)}}(\Phi(I))$ which is locally optimal for $\Phi(I)$, but is not locally optimal for $I$. By Lemma 16, $\mathsf{S}$ is position-consistent. This implies that $\Psi(I, \mathsf{S})$ is a legal assignment of values to variables $x \in \mathcal{X}$. Since by assumption, $\Psi(I, \mathsf{S})$ is not locally optimal for $I$, there exists a variable $x \in \mathcal{X}$ in instance $I \in$ POSNAE, which can be flipped such that clauses $C_s(x,y), \ldots, C_t(x,z) \in \mathcal{C}$ now have literals with non-identical assignments, clauses $C_p(x,u), \ldots, C_q(x,v) \in \mathcal{C} \in \mathcal{C}$ now have literals with identical assignments and the cost strictly increases by $z > 0$. This implies that in $\Phi(I)$, set $C_{x_i}^{IP} \in \mathsf{S}$ on position $\sigma(x)$ in $\mathsf{S}$ can be replaced by set $C_{x_{\bar{i}}}^{IP} \in \mathcal{C}$. Now, by construction $|C_{x_{\bar{i}}}^{IP} \cap C_{y_*}^{IP}| = \cdots = |C_{x_{\bar{i}}}^{IP} \cap C_{z_*}^{IP}| = 2 = a_{\sigma(x)\sigma(y)} = \cdots = a_{\sigma(x)\sigma(z)}$ and $|C_{x_{\bar{i}}}^{IP} \cap C_{u_*}^{IP}| = \cdots = |C_{x_{\bar{i}}}^{IP} \cap C_{v_*}^{IP}| = 0 \neq a_{\sigma(x)\sigma(u)} = \cdots = a_{\sigma(x)\sigma(v)}$. By definition of $B$, this strictly increases the cost of $\mathsf{S}$ by $z$. Thus, $\mathsf{S}$ is not locally optimal. A contradiction. □

### 3.9 On the Complexity of COMPARATIVECONTAINMENT-$(k)$

In this subsection, we prove that CC-$(k)$ is $\mathcal{PLS}$-complete for all $k \geq 1$. Given an instance $I \in (h)$-CNFSAT, we construct an instance $\Phi(I) = (\mathcal{M}, \mathcal{N}, w) \in D_{\text{CC-(1)}}$. Here, $\Phi(I)$, consists of two collections $\mathcal{M}$ and $\mathcal{N}$ of sets over a finite set $\mathcal{B}$ and a weight function $w : \mathcal{M} \cup \mathcal{N} \mapsto \mathbb{N}$ that maps sets from collections $\mathcal{M} \cup \mathcal{N}$ to positive integer weights.

*The Reduction* In a nutshell, the main idea is to *encode every satisfying assignment of a clause via sets* containing the respective literals and possessing the weight of the clause. This is polynomial in the size of the input since the length of a clause in $I$ is at most $h$. Additionally, we add *large incentives* to *exclude both literals of a variable* and *medium incentives* to *include literals*.

In more detail, we create an instance of CC-(1) over the finite set $\mathcal{B} := \{x, \bar{x} \mid x \in \mathcal{X}\}$. For every $x \in \mathcal{X}$, we introduce sets $X_x^{CC}$ in $\mathcal{N}$ with $w(X_x^{CC}) := 2W$, where $X_x^{CC} := \{y, \bar{y} \mid y \in \mathcal{X}; y \neq x\}$. This terminates the description of $\mathcal{N}$. Now, we define collection $\mathcal{M}$. Let $R_x := \{y, \bar{y} \mid y \in \mathcal{X}; y \neq x\}$. For every $x \in \mathcal{X}$, we introduce sets $C_x^{CC} := R_x \cup \{x\}$ and $C_{\bar{x}}^{CC} := R_x \cup \{\bar{x}\}$ in $\mathcal{N}$ with $w(C_x^{CC}) = w(C_{\bar{x}}^{CC}) := W$. Let $F_{C_i} := \{x, \bar{x} \mid x \text{ does not appear in } C_i\}$ denote the set of all fan-out variables in $\mathcal{B}$ which are irrelevant for satisfying $C_i$. Extending the technique from proof of Lemma 13, for every clause $C_i \in \mathcal{C}$, we encode every satisfying assignment for $C_i$ by the respective set, add all variables irrelevant for satisfying the clause, define it to possess the weight of clause $C_i$ and add it to $\mathcal{N}$. In detail, for every clause $C_i(x_{i_1}, \ldots, x_{i_k}) \in \mathcal{C}$ and every satisfying assignment $\varphi : \{x_{i_1}, \ldots, x_{i_k}\} \mapsto \{0,1\}^{i_k}$ for $C_i$, we introduce sets $C_x^{CC[\varphi(x_{i_1},\ldots,x_{i_k})]}$ in $\mathcal{M}$ with $w(C_i^{CC[\varphi(x_{i_1},\ldots,x_{i_k})]}) := w_{C_i}$. Here, $C_x^{CC[\varphi(x_{i_1},\ldots,x_{i_k})]} := F_{C_i} \cup \{\varphi'(x_{i_1}), \ldots, \varphi'(x_{i_j}), \ldots, \varphi'(x_{i_k})\}$ where $\varphi'(x_i) := x_i$ if $P_i(\varphi(x_{i_1}, \ldots, x_{i_k})) = 1$ and $\varphi'(x_i) := \bar{x}_i$ otherwise. The definition of an element-consistent solution $\mathsf{S} \in F_{\text{CC-(1)}}(\Phi(I))$ and the function $\Psi(I, \mathsf{S})$ is as in proof of Lemma 15.

**Lemma 18.** *Every locally optimal solution* $\mathsf{S} \in F_{\text{CC-(1)}}(\Phi(I))$ *is element-consistent.*

*Proof.* Assume there exists a solution $\mathsf{S}' \in F_{\text{CC-(1)}}(\Phi(I))$ with $|\mathsf{S}'| < |\mathcal{B}|/2$ and $\mathsf{S}'$ is locally optimal. Thus, there exists an element $x \in \mathcal{B}$ with $x \notin \mathsf{S}'$ and $\bar{x} \notin \mathsf{S}'$. Adding $x$ to $\mathsf{S}'$ increases

the cost of $S'$, since now $S' \not\subseteq X_x^{CC}$. The weight of $X_x^{CC}$ both dominates the sum of the weights lost due to $S' \not\subseteq C_{\bar{x}}^{CC}$ and the sum of the weights of the remaining sets in $\mathcal{M}$ in which $S'$ is not entirely contained any more. A contradiction.

Assume there exists a solution $S' \in F_{\text{CC-}(1)}(\Phi(I))$ with $|S'| > |\mathcal{B}|/2$ and $S'$ is locally optimal. By pigeonhole principle there exists an element $y \in S'$ with $\bar{y} \in S'$. Removing $\bar{y} \in S'$ increases the cost of $S'$, since, on the one hand, it does not alter any containment of $S'$ in $\mathcal{N}$. On the other hand, $S' \subseteq C_y^{CC}$ and the weight of $w(C_y^{CC}) = W$ dominates the sum of the smaller weights of sets in $\mathcal{M}$ in which $S'$ is not entirely contained any more. A contradiction.

Assume there exists a solution $S' \in F_{\text{CC-}(1)}(\Phi(I))$ with $|S'| = |\mathcal{B}|/2$ and $S'$ contains two elements $x \in S'$ and $\bar{x} \in S'$, and is locally optimal. By pigeonhole principle, there exists an element $y \in \mathcal{B}$ with $y \notin S'$ and $\bar{y} \notin S'$. Thus, exchanging $x$ for $y$ increases the cost of $S'$, since now $S' \not\subseteq X_y^{CC}$ and the weight of $X_y^{CC}$ dominates both the sum of the weight lost due to $S' \not\subseteq C_{\bar{x}}^{CC}$ and the sum of the remaining weights of sets in $\mathcal{M}$ in which $S'$ is not entirely contained any more. A contradiction. □

**Lemma 19.** $(h)$-CNFSAT $\leq_{pls}$ CC-$(k)$ *for all* $k \geq 1$.

*Proof.* Assume that there exists a feasible solution $S \in F_{\text{CC-}(1)}(\Phi(I))$ which is locally optimal for $\Phi(I)$, but is not locally optimal for $I$. By Lemma 18, $S$ is element-consistent. This implies that $\Psi(I, S)$ is a legal assignment for all variables $y \in \mathcal{X}$. Since $\Psi(I, S)$ is not locally optimal for $I$, there exists a variable $x \in \mathcal{X}$ in (h)-CNFSAT, which can be flipped such that clauses $C_i, \ldots, C_j \in \mathcal{C}$ become satisfied, clauses $C_p, \ldots, C_q \in \mathcal{C}$ become unsatisfied and the cost strictly increases by $z > 0$. This implies that in $\Phi(I)$, element $x \in S$ can be replaced by element $\bar{x} \in \mathcal{B}$. Now, $S$ is entirely contained in exactly one satisfying assignment encoded in some set $C_t^* \in \mathcal{M}$ from each family of sets $C_i^{CC[\varphi(x_{i_1}, \ldots, x_{i_k})]}, \ldots, C_j^{CC[\varphi(x_{j_1}, \ldots, x_{j_k})]} \in \mathcal{M}$, and $S$ is not entirely contained in all sets $C_p^{CC[\varphi(x_{p_1}, \ldots, x_{p_k})]}, \ldots, C_q^{CC[\varphi(x_{q_1}, \ldots, x_{q_k})]} \in \mathcal{M}$. By definition of $w$, this strictly increases the cost of $S$ by $z$. Thus, $S$ is not locally optimal. A contradiction. □

*Acknowledgement* The first author thanks Petra Berenbrink for many fruitful discussions.